\def\real{{\tt I\kern-.2em{R}}}
\def\nat{{\tt I\kern-.2em{N}}}
\def\hyper#1{\ ^*\kern-.2em{#1}} 
\def\hypernat{{^*{\nat }}}
\def\eskip{\hskip.25em\relax} 
\def\Hyper#1{\hyper {\eskip #1}}
\def\power#1{{{\cal P}(#1)}}
\def\iff{\leftrightarrow}

\def\m@th{\mathsurround=0pt}

\def\id{\par\hangindent2\parindent\textindent}
\def\textindent#1{\indent\llap{#1}}
\magnification=\magstep1
\tolerance 10000
\baselineskip  12pt
\hoffset=.25in
\hsize 6.00 true in
\vsize 8.85 true in
\font\eightrm=cmr9
\centerline{\bf An Ultimate Hyperfinite Ultralogic Unification}
\centerline{{\bf for Collections of Physical Theories}\footnote*{Any typographical errors that appear in the published version of this paper are caused by faulty publisher editing.}}\par\bigskip 
\centerline{Robert A. Herrmann}\par\medskip
\centerline{Mathematics Department}
\centerline{U. S. Naval Academy}
\centerline{572C Holloway Rd.}
\centerline{Annapolis,  MD 21402-5002}
\centerline{\ }\bigskip
{\leftskip=0.5in \rightskip=0.5in \noindent {\eightrm {\it Abstract:} Let $\{\rm S'_{{\rm N}_i}\mid i\in \nat\}$ represent a set of consequence operators defined on a language $\Lambda ,$ where each member of $\{\rm S'_{{\rm N}_i}\mid i\in \nat\}$ corresponds to a science-community physical theory and each $\rm N_i$ is a $\rm S'_{{\rm N}_j}$-system for each $\rm j \in \nat$. It is shown that there exists a hyperfinite ultralogic $\cal U \in \Hyper {{\cal C}_{\bf f}}$ defined on all internal subsets of $\hyper {{\bf \Lambda }}$ such that ${\cal U} \not= \Hyper {{\bf U}},$ and, for each $i,j \in \nat,\  \Hyper {{\bf S'_{\bf N}}_i}(\Hyper {{\bf N}}_j) = {\cal U}(\Hyper {{\bf N}_j}).$  For each  internal $Y \subset \hyper {{\bf \Lambda}},\  \bigcup\{\Hyper {{\bf S'_{\bf N}}_i}({ Y})\mid i \in \nat\} \subset {\cal U}({ Y})\subset \hyper {{\bf \Lambda}}.$ Further, if finite $\rm X \subset \Lambda,$ then $\bigcup\{
\Hyper {{\bf S'_{\bf N}}_i}({\bf X})\mid i \in \nat\} \subset {\cal U}({\bf X}),$  and if each member of $\rm \{ S'_{N_i}\mid i \in \nat\}$ is a practical consequence operator, then  $\bigcup\{
{\bf S'_{\bf N}}_i({\bf X})\mid i \in \nat\} \subset {\cal U}({\bf X}),$ and, for each $i,j \in \nat,\ {\bf S'_{ N}}_i({\bf N}_j) = {\cal U}( {\bf N}_j).$ Standard unifications for physical theories are also given.
 \par}}\par\bigskip
\noindent{\bf 1. Introduction.}\par\medskip
In Herrmann (2001a, b), a restricted hyperfinite ultralogic unification is constructed. The restrictions placed upon this construction were necessary in order to relate the constructed ultralogic directly to the types of ultralogics used to model probability models (Herrmann 2001c, d). In particular, the standard collections of consequence operators are restricted to a very special set of operators $\rm H_X$, where $\rm X$ is itself restricted to the set of all significant members of a language $\Lambda.$ In this paper, all such restrictions are removed. For reader convince, some of the introductory remarks that appear in Herrmann (2001a, b) are repeated. Over seventy years ago, Tarski (1956, pp. 60-109) introduced {\it consequence} operators as models for various aspects of human thought. There are two such mathematical theories investigated, the {\it general} and the {\it finitary} consequence operators (Herrmann, 1987). Let $\rm L$ be a nonempty language, $\cal P$ be the power set operator and $\cal F$ the finite power set operator. There are three cardinality independent axioms.\par\medskip
{\bf Definition 1.1.} A mapping $\rm C\colon \power{\rm L} \to \power {\rm L}$ is a general consequence operator (or closure operator) if for each $\rm X,\ \rm Y \in 
\power {\rm L}$\par\smallskip
\indent\indent (1) $\rm X \subset C(X) = C(C(X)) \subset L$; and if\par\smallskip
\indent\indent (2) $\rm X \subset Y$, then $\rm C(X) \subset C(Y).$\par\smallskip\vfil\eject
\noindent A consequence operator C defined on L is said to be {\it finitary} ({\it finite}) if it satisfies\par\smallskip
\indent\indent (3) $\rm C(X) = \bigcup\{C(A)\mid A \in {\cal F}(\rm X)\}.$\par\medskip
{\bf Remark 1.2.} The above axioms (1), (2), (3) are not independent. Indeed, 
(1) and (3) imply (2). Clearly, the set of all finitary consequence operators defined on a specific language is a subset of the set of all general operators. The phrase ``defined on $\rm L$'' means formally defined on $\power {\rm L}.$ \par\medskip
All known scientific logic-systems use finitely many rules of inference and finitely many steps in the construction of a deduction from these rules. Hence, as shown in  Herrmann (2001a, b), the consequence operator that models such theory generating thought processes is a finitary consequence operator. Although many of the results in this paper hold for the general consequence operator, we are only interested in collections of finitary consequence operators. Dziobiak (1981, p. 180) states the Theorem 2.10 below. However, the statement is made without a formal proof and is relative to a special propositional language. Theorem 2.10 is obtained by using only basic set-theoretic notions and Tarski's basic results for any language. Further, the proof reveals some interesting facts not previously known. Unless noted, all utilized Tarski (1956, pp. 60-91) results are cardinality independent.\par\medskip
\noindent {\bf 2. The Lattice of Finitary Operators.}\par\medskip
 {\bf Definition 2.1.} In all that follows, any set of consequence operators will be nonempty and each is defined on a nonempty language. Define the relation $\leq$ on the set $\rm {\cal C}$ of all general consequence operators defined on $\rm L$ by stipulating that for any $\rm C_1, C_2 \in {\cal C},$ $\rm C_1 \leq C_2$ if for every $\rm X \in \power {\rm L},\ 
C_1(X) \subset C_2(X).$ \par\medskip
\noindent Obviously, $\leq$ is a partial order contained in $\rm {\cal C}\times {\cal C}.$ Our standard result will show that for the entire set of finitary consequence operators $\rm {\cal C}_f \subset {\cal C}$ defined on $\rm L$, the structure $\langle {\cal C}_f,\leq \rangle$ is a lattice. \par\medskip
{\bf Definition 2.2.} Define $\rm I \colon \power {L} \to \power {L}$ and $\rm U \colon \power {L} \to \power {L}$ as follows: for each $\rm X\subset L,$ let $\rm I(X) = X,$ and let $\rm U(X) = L.$ \par\medskip 
\noindent Notice that $\rm I$ is the lower unit (the least element) and $\rm U$ the upper unit (the greatest element) for $\rm \langle {\cal C}_f,\leq \rangle$ and $\rm \langle {\cal C},\leq \rangle.$\par\medskip
{\bf Definition 2.3.} Let $\rm C \in {\cal C}.$ A set $\rm X \subset {\rm L}$ is a 
$\rm C$-system or simply a system if $\rm C(X) \subset X$ and, hence, if $\rm C(X) = X.$ For each $\rm C \in {\cal C},$ let $\rm {\cal S}(C) = \{ X\mid (X \subset {\rm L})\land (C(X) = X)\}.$ \par\medskip
\noindent Since $\rm C(L) = L$ for each $\rm C \in {\cal C},$ then each $\rm {\cal S}(C) \not= \emptyset.$\par\medskip
{\bf Lemma 2.4} {\it For each $\rm C_1,\ C_2 \in {\cal C},$ $\rm C_1 \leq C_2$ if and only if $\rm {\cal S}(C_2) \subset {\cal S}(C_1).$}\par\smallskip 
Proof. Let any $\rm C_1,\ C_2 \in {\cal C}$ and $\rm C_1 \leq C_2.$ Consider any $\rm Y\in {\cal S}(C_2).$ Then $\rm C_1(Y) \subset C_2(Y)= Y.$ Thus, $\rm C_1 \in {\cal S}(C_1)$ implies that $\rm {\cal S}(C_2) \subset {\cal S}(C_1).$\par\smallskip 
Conversely, suppose that $\rm {\cal S}(C_2) \subset {\cal S}(C_1).$
 Let $\rm X \subset L.$ Then since, by axiom 1, $\rm C_2(X) \in {\cal S}(C_2)$, it follows, from the requirement that $\rm C_2(X) \in {\cal S}(C_1),$ that 
$\rm C_1(C_2(X)) = C_2(X).$ But $\rm X\subset C_2(X)$ implies that $\rm C_1(X) \subset C_1(C_2(X))= C_2(X),$ from axiom 2. Hence, $\rm C_1 \leq C_2$ and the proof is complete.\par\medskip
{\bf Definition 2.5.} For each $\rm C_1,\ C_2 \in {\cal C}$, define the following binary relations in $\rm \power {L} \times \power {L}$. For each $\rm X \subset L,$ let $\rm (C_1 \land C_2)(X) = C_1(X) \cap C_2(X)$ and $\rm (C_1 \lor_w C_2) = \bigcap \{Y\subset L\mid (X \subset Y = C_1(Y)=C_2(Y))\}$  For finitely many members of $\cal C,$ the operators $\land,\ \lor_w$ are obviously commutative and associative. These two relations are extended to arbitrary ${\cal A} \subset {\cal C}$ by defining $\rm (\bigwedge {\cal A})(X) = \bigwedge {\cal A}(X)=\bigcap \{C(X)\mid C \in {\cal A}\}$ and $\rm (\bigvee_w {\cal A})(X) = \bigvee_w {\cal A}(X) =\bigcap \{Y \subset L\mid X\subset Y = C(Y)\ {\rm for\ all}\ C \in {\cal A}\}$ (Dziobiak, 1981, p. 178). Notice that $\rm \bigvee_w {\cal A}(X) =\bigcap \{Y \subset L\mid (X\subset Y)\land (Y \in \bigcap \{{\cal S}(C)\mid C \in {\cal A}\})\}.$\par\medskip
{\bf Lemma 2.6.} {\it Let ${\cal A} \subset {\cal C}$ {\rm [}resp. $\rm {\cal C}_f${\rm ]} and $\rm {\cal S}'  = \{ X\mid (X \subset L)\land (X = \bigvee_w {\cal A}(X))\}.$ Then $\rm {\cal S}' = \bigcap \{{\cal S}(C)\mid C \in {\cal A}\}.$} \par\smallskip
Proof. By Tarski's Theorem 11 (b) (1956, p. 71), which holds for finitary and general consequence operators, for each $\rm X \subset L$ and $\rm C \in {\cal A},\ X\subset \bigvee_w {\cal A}(X) = Y' \in {\cal S}(C).$ Hence, if $\rm Y' \in {\cal S}',$ then $\rm \bigvee_w {\cal A}(Y') = Y'\in {\cal S}(C)$ for each $\rm C \in {\cal A}.$ Thus $\rm {\cal S}' \subset \bigcap \{{\cal S}(C)\mid C \in {\cal A}\}.$ Conversely, let $\rm Y \in  \bigcap \{{\cal S}(C)\mid (C \in {\cal A})\}.$ From the definition of $\rm \bigvee_w,\ \bigvee_w {\cal A}(Y) = Y$ and, hence, $\rm Y \in {\cal S}'$ and this completes the proof.\par\medskip
{\bf Lemma 2.7.} {\it Let nonempty ${\cal B} \subset \rm \power {L}$ and $\rm L \in {\cal B}.$ Then the operator $\rm C_{\cal B}$ defined for each $\rm X \subset L$ by $\rm C_{\cal B}(X) = \bigcap \{Y\mid X \subset Y \in {\cal B}\}$ is a general consequence operator defined on $\rm L.$} \par\smallskip
Proof. Assuming the hypothesis, it is obvious that $\rm C_{\cal B} \colon {\cal P}(L) \to {\cal P}(L)$ and $\rm X \subset C_{\cal B}(X).$ Clearly, if $\rm Z \subset X \subset L$, then $\rm C_{\cal B}(Z) \subset C_{\cal B}(X);$ and, for each $\rm Y \in {\cal B},$ $\rm X \subset Y$ if and only if $\rm C_{\cal B}(X) \subset Y.$ Hence, $\rm C_{\cal B}(C_{\cal B}(X)) = \bigcap\{Y\mid C_{\cal B}(X) \subset Y \in {\cal B}\} = C_{\cal B}(X).$  This completes the proof.\par\medskip
{\bf Remark 2.8.} The hypotheses of Lemma 2.7 do not require that the set $\cal B$ be closed under arbitrary intersection. \par\medskip
{\bf Theorem 2.9.} {\it With respect to the partial order relation $\leq$  defined on $\rm L,$ the structure $\rm \langle {\cal C}, \lor_w, \land, I, U \rangle$ is a complete lattice with upper and lower units.}\par\smallskip
Proof. 
Let ${\cal A} \subset {\cal C}$ and $\rm {\cal B} = \bigcap \{{\cal S}(C)\mid C \in {\cal A}\}.$ Since $\rm L \in {\cal B},$ then by Lemma 2.7, $\rm \bigvee_w {\cal A}= C_{\cal B} \in {\cal C}.$ Moreover, by Lemmas 2.4 and 2.6, $\rm C_{\cal B}$ is the least upper bound for  $\cal A$ with respect to $\leq.$\par\smallskip
Next, let  $\rm {\cal B} = \bigcup \{{\cal S}(C)\mid C \in {\cal A}\}$. For $\rm X \subset L,\ X \subset C(X)$ for each $\rm C \in {\cal A}.$ For each $\rm C \in {\cal A},$ there does not exist a $\rm Y_C$ such that $\rm Y_C \in {\cal S}(C),\ X \not= Y_C, Y_C \not= C(X)$ and $\rm X \subset Y_C \subset C(X).$ Hence, $\rm C_{\cal B}(X) = \bigcap\{Y\mid X \subset Y \in {\cal B}\} = \bigcap \{C(X) \mid C \in {\cal A}\} = \bigwedge {\cal A}(X).$ Hence, $\rm \bigwedge {\cal A} \in {\cal C}$ and it is obvious that $\rm \bigwedge {\cal A}$ is the greatest lower bound for $\cal A$ with respect to $\leq.$ This completes the proof.\par\medskip
Although the proof appears in error, (W\'ojcicki, 1970) stated Theorem 2.9  for a propositional language. In what follows, we only investigate the basic lattice structure for $\rm \langle {\cal C}_f, \leq \rangle .$ \par\vfil\eject
{\bf Theorem 2.10.} {\it With respect to the partial order relation $\leq$ defined on $\rm {\cal C}_f,$ the structure $ \langle {\cal C}_f, \lor_w, \land, I, U \rangle$ is a lattice with upper and lower units.}\par\smallskip
Proof.  It is only necessary to consider two distinct $\rm C_1,\ C_2 \in {\cal C}_f.$ As mentioned, the commutative and associative laws hold for $\land$ and $\lor_w$ and by definition each maps $\rm \power {L}$ into $\rm \power {L}$. In $\langle {\cal C},\leq \rangle$, using theorem 2.9, axiom 1 and 2 hold for the greatest lower bound $\rm C_1 \land C_2$ and for the least upper bound $\rm C_1 \lor_w C_2.$ Next, we have that $\rm (C_1 \land C_2)(X) = (\bigcup \{C_1(Y) \mid Y \in {\cal F}(X)\}) \cap (\bigcup \{ C_2(Y) \mid Y \in {\cal F}(X)\})=\bigcup \{C_1(Y) \cap C_2(Y)\mid Y \in {\cal F}(X)\} = \bigcup \{(C_1 \land C_2)(Y)\mid Y \in {\cal F}(X)\}$ and axiom 3 holds and, hence, $\rm C_1 \land C_2 \in {\cal C}_f.$ Therefore, $\rm \langle {\cal C}_f,\land, I, U \rangle$ is, at the least, a meet semi-lattice.\par\smallskip

Next, we show by direct means that for each $\rm C_1,\ C_2 \in {\cal C}_f,$ $\rm C_1 \lor_w C_2 \in {\cal C}_f.$  Let (the cardinality of L) $\rm \vert L\vert = \Delta.$ For each $\rm X_i \subset L,\ (i \in \Delta),$ let $\rm {\cal A}'(X_i)= \{Y\mid (X_i\subset Y \in {\cal S}(C_1)\cap {\cal S}(C_2) )\land (Y \subset L)\}.$   Let $\rm \bigcap \{Y\mid Y \in {\cal A}'(X_i)\} = Y_i.$ By Tarski's  Theorem 11a (1956, p. 71) $\rm X_i \subset Y_i \in {\cal S}(C_1) \cap {\cal S}(C_2),$ and by definition $ \rm Y_i = (C_1 \lor_w C_2)(X_i).$ Hence, $\rm Y_i \in {\cal A}'(X_i)$ and is the least ($\subset$) element.  
For $\rm X_i \subset L,$ let $\rm {\cal A}''(X_i)= \{Y\mid (C_1(X_i)\subset Y \in {\cal S}(C_1) \cap {\cal S}(C_2))\land (Y \subset L)\}.$ Since  $\rm X_i \subset C_k(X_i),\ k = 1, 2$, then $\rm {\cal A}'' \subset {\cal A}'.$ 
Since $\rm L \in {\cal A}'(X_i),\ {\cal A}'(X_i)\not=\emptyset.$ Indeed, let
$\rm Y \in {\cal A}'(X_i).$ Then $\rm X_i \subset C_k(Y) = Y, \  k = 1,2.$ Additionally, $\rm X_i \subset C_1(Y)=Y$ implies that $\rm X_i \subset  C_1(X_i)=C_1(C_1(X_i))\subset C_1(C_1(Y))=C_1(Y) = Y.$ Hence,  
it follows that for any $\rm X_i \subset L,\ {\cal A}''(X_i) =  {\cal A}'(X_i).$ For fixed $\rm X_i \subset L,$ let $\rm X_j \in {\cal F}(X_i).$ Let $\rm Y_j$ be defined as above and, hence, $\rm Y_j$ is the least element in $\rm {\cal A}'(X_j)= {\cal A}''(X_j).$ Consider $\rm {\cal D}= \{Y_j\mid X_j \in {\cal F}(X_i)\},$ and, for $\rm j = 1,\ldots,n,$ consider $\rm Y_j \in {\cal D}$ and the corresponding $\rm X_j \subset L.$ 
Let $\rm X_k = \bigcup \{X_j\mid j = 1,\ldots, n\}\in {\cal F}(X_i).$ Then $\rm Y_k = \bigcap\{Y \mid Y \in {\cal A}'(X_k)\}\in {\cal D}.$ If $\rm Y \in {\cal A}'(X_k),$ then $\rm Y \in {\cal A}'(X_j),\ j = 1,\ldots,n.$ Hence, $\rm Y_j \subset Y_k, \ j = 1,\ldots,n$
implies that $\rm Y_1 \cup \cdots \cup Y_n \subset Y_k.$ Tarski's Theorem 12 (1956, p. 71) implies that $\rm Y^* = \bigcup \{Y_j \mid X_j \in {\cal F}(X_i)\} \in {\cal S}(C_1)\cap {\cal S}(C_2).$ Also, by definition, for all $\rm X_j \subset L,$ $\rm Y_j \in {\cal A}''(X_j)$ implies that $\rm C_1(X_j) \subset Y_j.$ The fact that $\rm C_1$ is finitary yields $\rm C_1(X_i) \subset Y^*.$ Hence, $\rm Y^*\in {\cal A}''(X_i).$ Since $\rm C_1(X_j) \subset C_1(X_i),\ X_j \in {\cal F}(X_i),$ then $\rm {\cal A}''(X_i) \subset {\cal A}''(X_j).$ Thus $\rm Y_j \subset Y_i,\ X_j \in {\cal F}(X_i).$ Therefore, $\rm Y^* \subset Y_i.$ But, $\rm Y^* \in {\cal A}''(X_i)$ implies that $\rm Y^* = Y_i.$ Re-stating this last result, $\rm \bigcup \{ (C_1\lor_w C_2)(X_j)\mid X_j \in {\cal F}(X_i)\} = (C_1\lor_w C_2)(X_i)$ and, therefore, axiom (3) holds for the binary relation $\lor_w$ and $\rm \langle {\cal C}_f, \lor_w, \land, I, U \rangle$ is a lattice. This completes the proof.\par\medskip
{\bf Corollary 2.10.1.} {\it For the set $\rm {\cal C}_f$ of all finitary consequence operators defined on $\rm L,$ the structure $\rm \langle {\cal C}_f, \wedge,\lor_w, I, U \rangle$ is a join-complete lattice.}\par\smallskip
Proof. Let non-empty $\rm {\cal A} \subset {\cal C}_f.$ Now simply modify the second part of the proof of Theorem 2.10 by substituting $\rm \bigcap \{{\cal S}(C)\mid C \in {\cal A}\}$ for $\rm {\cal S}(C_1) \cap {\cal S}(C_2)$ and letting $\rm C_1 \in {\cal A}.$ This complete the proof.\par\medskip 
{\bf Remark 2.11.}  Tarshi's Theorem 12 requires finitary consequence operators.  It is known, since $\rm I$ is a lower bound for any $\rm {\cal A} \subset {\cal C}_f$ that $\rm \langle {\cal C}_f, \lor_w, I, U \rangle$ is actually a complete lattice with a meet operator generated by the $\bigvee_w$-operator. It appears that this meet operator need not correspond, in general, to results for the defined $\bigwedge {\cal A}$ operator for an infinite $\cal A.$ W\'ojcicki (1973) has constructed, for a set of consequnece operators ${\cal C}',$ an infinite $\rm {\cal A}\subset {\cal C}'$ of finitary consequence operators,  with some very special properties. However, the general consequence operator defined for each $\rm X \subset L$ by $\rm \bigcap \{C(X)\mid C \in {\cal A}\}$ is not a finitary operator. Thus, in general, $\rm \langle {\cal C}_f, \lor_w, \land,I,U \rangle$ need not be a complete sublattice of  $\rm \langle {\cal C}, \lor_w, \land,I,U \rangle.$ This behavior is not unusual. For example, let infinite $\rm X$ have an infinite topology $\cal T.$ Then $\langle {\cal T},\cup,\emptyset, X \rangle$ is a join-complete subsemi-lattice of the complete lattice $\langle \power{X},\cup, \cap,\emptyset, X\rangle.$ The structure $\langle {\cal T},\cup,\emptyset, X \rangle$is actually complete, but it is not a sublattice of $\langle \power{X},\cup, \cap,\emptyset, X\rangle.$ \par\medskip  
\noindent {\bf 3. System Consistent Logic-systems} \medskip
Let $\Sigma$ be a non-empty set of science-community logic-systems and let $\vert \cdot\vert$ denote cardinality. In practice, $\vert \Sigma\vert \leq \aleph_0.$ Each logic-system $\rm S_i \in \Sigma,\ i \in \vert\Sigma\vert,$ is defined on a countable language $\rm L_i$ and each $\rm S_i$ determines a specific finitary consequence operator $\rm C_i$ defined on a language $\rm L_i.$ At the least, by application of the insertion of hypotheses rule (Herrmann, 2001a/b, p. 94/2), for nonempty cardinal $\Delta \leq \vert \Sigma\vert,$ each member of $\rm \{C_i\mid i \in \Delta\}$ is defined on the language $\rm \bigcup\{L_i\mid i \in \Delta\}.$  {\it In all that follows, a specific set of logic-system generated consequence operators $\rm \{C_i\mid i \in \Delta\}$ defined on a specific set of languages $\rm \{L_i \mid i \in \Delta\}$ will always be considered as trivially extended and, hence, defined by the insertion of hypotheses rule on the set $\rm \bigcup\{L_i\mid i \in \Delta\}$.} In general, such a specific set of consequence operators is contained in the lattice of all finitary operators defined on $\rm \bigcup\{L_i\mid i \in \Delta\}$. A logic-system $\rm S'$ and its corresponding consequence operator is a {\it trivial extension} of a logic-system's $\rm S$ defined on $\rm L$ where, for a language $\rm L' \supset L,$ $\rm S'$ is the same as $\rm S$ except that only the hypotheses insertion rule now applies to $\rm L'-L$. The system $\rm S'$ and its corresponding consequence operator $\rm C'$ is a non-trivial extension if it is extended to $\rm L'$ by insertion and some other n-ary relations that contain members of $\rm L' -L$ are adjoined to those in $\rm S$ or various original n-ary relations in $\rm S$ are extended by adding n-tuples that contain members from $\rm L' - L.$  For both the trivial and non-trivial cases and with respect to the language $\rm L'$, it follows that $\rm C \leq C'.$ In the trivial case, if $\rm X \subset L',$ then $\rm C'(X) = C(X \cap L) \cup (X-L).$ \par\smallskip
In practice, a {\it practical logic-system} is a logic-system defined for the subsets of a finite language $\rm L^f.$ When a specific deduction is made from a set of hypotheses $\rm X,$ the set $\rm X$ is finite. If the logic-system also includes 1-ary sets, such as the logical or physical axioms, the actual set of axioms that might be used for a deduction is also finite. Indeed, the actual set of all deductions obtained at any moment in human history and used by a science-community form a finite set of statements that are contained in a finite language $\rm L^f$. (Finite languages, the associated consequence operators and the like will usually be denoted by a $\rm f$ superscript.) The finitely many n-ary relations that model the rules of inference for a practical logic-system are finite sets.  Practical logic-systems generate practical consequence operators and  practical consequence operators  generate effectively practical logic-systems, in many ways. For example, the method found in \L o\'s, J. and R. Suszko (1958), when applied to a $\rm C^f,$ will  generate effectively a finite set of rules of inference. The practical logic-system obtained from such rules generates the original practical consequence operator. Hence, a consequence operator $\rm C^f$ defined on $\rm L^f$ is considered a {\it practical consequence operator} although it may not correspond to a previously defined scientific practical logic-system; nevertheless, it does correspond to an equivalent practical logic-system.\par\smallskip
Our definition of a {\it physical theory} is a refinement of the usual definition. Given a set of physical hypotheses, general scientific statements are deduced. If accepted by a science-community, these statements become {\it natural laws.} These natural laws then become part of a science-community's logic-system. In Herrmann (2001a, b, a), a consequence operator generated by such a logic-system is denoted by $\rm S_N.$ From collections of such logic-systems, the $\rm S_N$ they generate are then applied to specific natural-system descriptions $\rm X.$ For scientific practical logic-systems, the language and rules of inference need not be completely determinate in that, in practice, the language and rules of inference are extended.  \par\smallskip

The complete Tarski definition for a consequence operator includes finite languages (1956, p. 63) and all of the Tarski results used in this paper apply to such finite languages. Theorem 2.10 holds for any language finite or not. In the lattice of finitary consequence operators defined on ${\rm L^f},$ $\lor_w$ determines the least upper bound for a finite set of such operators. However, it is certainly possible that this least upper bound is the upper unit $\rm U.$\par\medskip
{\bf Definition 3.1.} Let $\rm C$ be a general consequence operator defined in $\rm L.$ Let $\rm X\subset L.$ \par\smallskip
(i) The set $\rm X$ is C-{\it consistent} if $\rm C(X) \not= L.$\par\smallskip
(ii) The set $\rm X$ is C-{\it complete} if for each $\rm x \in L,$ either $\rm x \in X$ or $\rm C(X \cup \{x\})= L.$\par\smallskip
(iii) A set $\rm X \subset L$ is {\it maximally {\rm C}-consistent} if $\rm X$ is C-consistent and whenever a set $\rm Y \not = X$ and $\rm X \subset Y \subset L,$ then $\rm C(Y) = L.$\par\smallskip
\noindent Notice that if $\rm X \subset L$ is C-consistent, then $\rm C(X)$ is a C-consistent extension of $\rm X$ which is also a C-system. Further, C-consistent $\rm W$ is C-consistent with respect to any trivial extension of $\rm C$ to a language $\rm L' \supset L.$\par\medskip 
{\bf Theorem 3.2} {\it Let general consequence operator $\rm C$ be defined on $\rm L$.}\par\smallskip 
(i) {\it The set $\rm X \subset L$ is $\rm C$-complete and $\rm C$-consistent if and only if $\rm X$ is a maximally ${\rm C}$-consistent.}\par\smallskip
(ii) {\it If $\rm X$ is maximally $\rm C$-consistent, then $\rm X$ is a $\rm C$-system.}\par\smallskip
Proof. (i) Let $\rm X$ be maximally $\rm C$-consistent. Then $\rm X$ is C-consistent and, hence, $\rm C(X) \not= L.$ Hence, let $\rm x \in L$ and $\rm x \notin X.$ Then $\rm X \subset X \cup \{x\}$ implies that $\rm X \cup \{x\}$ is not $\rm C$-consistent. Thus $\rm C(X \cup \{x\})= L$. Hence, $\rm X$ is C-complete. Conversely, assume that $\rm X$ is C-consistent and C-complete. Then $\rm X\not= L.$ Let $\rm X\subset Y \subset L$ and $\rm X \not= Y.$ Hence, there is some $\rm y \in Y-X$ and from C-completeness $\rm L= C(X \cup \{y\}) \subset C(Y).$ Thus, $\rm Y$ is not C-consistent. Hence, $\rm X$ is maximally C-consistent and the result follows.\par\smallskip
(ii) From C-consistency, $\rm C(X) \not= L$. If $\rm x \in C(X) - X,$ then maximally C-consistent implies that $\rm L = C(X\cup \{x\})\subset C(C(X)) = C(X)$. This contradiction yields that $\rm X$ is a C-system.  \par\medskip 
The following easily obtained result holds for many types of languages (Tarski, 1956, p. 98. Mendelson, 1979, p. 66) but these ``Lindenbaum'' constructions, for infinite languages, are not considered as effective. For finite languages, such constructions are obviously effective.\par\medskip

{\bf Theorem 3.3.} {\it Let practical consequence operator $\rm C^f$ be defined on arbitrary $\rm L^f.$ If $\rm X \subset L^f$ is $\rm C^f$-consistent, then there exists an effectively constructed $\rm Y \subset L^f$ such that $\rm C^f(X)\subset Y$, $\rm Y$ is $\rm C^f$-consistent and $\rm C^f$-complete.}\par\smallskip 
Proof. This is rather trivial for a practical consequence operator and all of the construction processes are effective.  Consider an enumeration for $\rm L^f$ such that $\rm L^f = \{x_1, x_2,\ldots x_k\}.$  Let $\rm X \subset L^f$ be $\rm C^f$-consistent and define $\rm X = X_0.$ We now simply construct in a completely effective manner a partial sequence of subsets of $\rm L^f.$ Simply consider $\rm X_0 \cup \{x_1\}.$ Since $\rm X_0$ is $\rm C^f$-consistent, we have two possibilities. Effectively determine whether $\rm C^f(X_0 \cup \{x_1\}) = L^f.$ If so, let $\rm X_1 = X_0.$ On the other hand, if $\rm C^f(X_0 \cup \{x_1\}) \not= L^f,$ then define $\rm X_1 = X_0 \cup \{x_1\}.$ Repeat this construction finitely many times. (Usually, if the language is denumerable, this is expressed in an induction format.) Let $\rm Y = X_k.$ By definition, $\rm Y$ is $\rm C^f$-consistent. Suppose that $\rm x\in L^f.$ Then there is some $\rm X_i$ such that either (a) $\rm x \in X_i$ or (b) $\rm C^f(X_i \cup \{x\}) = L^f$.  For (a), since $\rm X_i \subset Y$, $\rm x \in Y.$ For (b), $\rm X_i \subset Y,$ implies that $\rm L = C^f(X_i \cup \{x\})\subset C^f(Y \cup \{x\}) = L^f.$ Hence, $\rm Y$ is $\rm C^f$-complete and $\rm X_i \subset Y,$ for each $\rm i = 1,\ldots, k.$ By Theorem 3.2, $\rm Y$ is a $\rm C^f$-system. Thus $\rm X_0 \subset Y$ implies that $\rm C^f(X_0) \subset C^f(Y) = Y,$ and this completes the proof.\par\smallskip
{\bf Corollary 3.3.1.} {\it Let practical consequence operator $\rm C^f$ be defined on $\rm L^f$ and $\rm X \subset L^f$ be $\rm C^f$-consistent. Then there exists an effectively constructed $\rm Y \subset L^f$ that is an extension of $\rm C^f(X)$ and, hence, also an extension of $\rm X,$ where $\rm Y$ is a maximally $\rm C^f$-consistent $\rm C^f$-system.}\par\medskip 
Let the set $\rm \Sigma^p \subset \Sigma$ consist of all of science-community practical logic-systems defined on languages $\rm L^f_i.$ Each member of $\rm \Sigma^p$ corresponds to $ \rm i \in \vert \Sigma^p \vert$ and to a practical consequence operator $\rm C_i^f$ defined on $\rm L^f_i.$ In general, the members of a set of science-community logic-systems are related by a consistency notion relative to an extended language. \par\medskip
{\bf Definition 3.4.} A set of consequence operators $\cal C$ defined on $\rm L$ is {\it system consistent} if there exists a $\rm Y \subset L,\ Y \not=L$ and $\rm Y$ is a C-system for each $\rm C \in \cal C.$   
\par\medskip\
{\bf Example 3.5.} Let $\cal C$ be a set of axiomless consequence operators where each $\rm C \in \cal C$ is define on $\rm L$. In Herrmann (2001a, b), the set of science-community consequence operators is redefined by relativization to produce a set of axiomless consequence operators, the $\rm S^V_N$, each defined on the same language. Any such collection $\cal C$ is system consistent since for each $\rm C \in {\cal C}, \ C(\emptyset) = \emptyset \not= L.$  \par\medskip
{\bf Example 3.6.} One of the major goals of certain science-communities is to find what is called a ``grand unification theory.'' This is actually a theory that will unify only the four fundamental interactions (forces). It is then claimed that this will somehow lead to a unification of all physical theories. Undoubtedly, if this type of grand unification is achieved, all other physical science theories will require some type of re-structuring. The simplest way this can be done is to use informally the logic-system expansion technique. This will lead to associated consequence operators defined on ``larger'' language sets.\par\smallskip
 Let a practical logic-system $\rm S_0,$ be defined on $\rm L^f_0,$ and $\rm L = \bigcup\{L^f_i\mid i \in \nat\},$ $\nat$ the set of natural numbers.  Let $\rm L_0 \subset L_1, L_0 \not= L_1.$ [Note: the remaining members of $\rm \{L^f_i\mid i \in \nat\}$ need not be distinct.] Expand $\rm S_0$ to $\rm S_1 \not= S_0$ defined on $\rm L$ by adjoining to the logic-system $\rm S_0$ finitely many practical logic-system n-ary relations or finitely many additional n-tuples to the original $\rm S_0,$ but  
where all of these additions only contain members from nonempty $\rm L - L^f_0.$ Although $\rm S_1$ need only be considered as non-trivially defined on $\rm L^f_1,$ if $\rm L\not= L_1,$ then the $\rm S_1$ so obtained corresponds to $\rm C_1,$ a consequence operator trivially extended to $\rm L.$ This process can be repeated in order to produce, at the least, finitely many distinct logic-systems $\rm S_i, \ i >1,$ that extend $\rm S_0$ and a set $\rm {\cal C}_1$ of distinct corresponding consequence operators $\rm C_i.$ Since these are science-community logic-systems, there is an $\rm X_0 \subset L^f_0$ that is $\rm C^f_0$-consistent. By Corollary 3.3.1, there is an effectively defined set $\rm Y \subset L^f_0$ such that $\rm X_0 \subset Y$ and $\rm Y$ is maximally $\rm C^f_0$-consistent with respect to the language $\rm L^f_0.$ Hence, $\rm C^f_0(Y) = Y \subset L^f_0$ and $\rm C^f_0(Y) \not= L^f_0.$ Further, $\rm C^f_0$ is consider trivially extended to $\rm L.$ Let $\rm Y' = Y \cup (L - L^f_0).$ It follows that for each $\rm C_i,\ L- L^f_0 \subset C_i(L - L^f_0) \subset L - L^f_0 \not= L.$ By construction, for each $\rm C_i,\ C_i(Y) = Y;$ and for each $\rm X \subset L,\ \rm C_i(X) = C_0(X \cap L^f_0) \cup C_i(X \cap (L - L^f_0)).$ So, let $\rm X = Y'$. Then for each $\rm C_i,\ C_i(Y') = C_0(Y) \cup (L -L^f_0)= Y \cup (L-L^f_0)= Y' \not= L.$  Hence, the set of all $\rm C_i$ is system consistent.\par\medskip
{\bf Example 3.7.} Consider a denumerable language $\rm L$ and Example 3.2 in Herrmann (1987). [Note: There is a typographical error in this 1987 example. The expression $x \notin {\cal U}$ should read $x \notin U.$] Let $\cal U$ be a free-ultrafilter on $\rm L$ and let $\rm x \in L.$ Then there exists some $\rm U \in {\cal U}$ such that $\rm x \notin U$ since $\bigcap {\cal U} = \emptyset$ and $\emptyset \notin {\cal U}.$ Let $\rm B = \{x\}$ and $\rm {\cal C} = \{P(U,B)\mid U \in {\cal U}\},$ where $\rm P(U,B)$ is the finitary consequence operator defined by $\rm P(U,B)(X) = U \cup X,$ if $\rm x \in X$;  and $\rm P(U,B)(X) = X,$ if $\rm x \notin X$. [Note: this is the same operator $\rm P$ that appears in the proof of Theorem 6.4 in Herrmann (2001a, b).] There, at the least, exists a sequence $\rm S = \{ U_i \mid i \in \nat\}$ such that $\rm U_0 = U$ and $\rm U_{i+1} \subset U_i,\ U_{i+1} \not= U_i.$ It  follows immediately from the definition that $\rm P(U_{i+1},B) \leq P(U_i,B)$ and $\rm P(U_{i+1},B)(B)= U_{i+1} \cup B \subset U_i \cup B,$ for each $\rm i \in \nat.$ Hence, in general, $\rm P(U_{i + 1}, B ) < P(U_i,B)$ for each $\rm i \in \nat.$  Let $\rm Y = L- \{x\}.$ Then $\rm P(U_i,B)(Y) = U_i \cup (L - \{x\}) = L-\{x\}=Y,\ i \in \nat.$ Thus, the collection $\rm \{P(U_i,B)\mid i\in \nat\}$ is system consistent.  \par\medskip 
{\bf Theorem 3.8.} {\it Consider $\rm {\cal A}\subset {\cal C}_f$ defined on $\rm L$ and the $(\leq)$ least upper bound $\rm \bigvee_w {\cal A}.$ Then $\rm \bigvee_w {\cal A}\in {\cal C}_f$ and if $\cal A$ is system consistent, then there exists some $\rm Y \subset L$ such that $\rm Y = \bigvee_w {\cal A}(Y) = C(Y) \not= L$ for each $\rm C \in {\cal A}$ and $\rm \bigvee_{ w}{\cal A} \not= U.$ Further, if $\rm X\subset L, X \not= L,$ is a ${\rm C}$-system for each $\rm C \in {\cal A},$ then $\rm X = \bigvee_w {\cal A}(X) = C(X) \not= L$ for each $\rm C \in {\cal A}.$}\par\smallskip   
Proof. Corollary 2.10.1 yields the first conclusion. From the definition of system consistent, there exists some $\rm Y \subset L$ such that  $\rm C(Y) = Y\not= L$ for each $\rm C \in {\cal A}.$ 
 From Lemma 2.6, for each $\rm C \in {\cal A},\ \bigvee_w {\cal A}(Y)= C(Y) \not= L.$ Hence, $\rm \bigvee_{w}{\cal A} \not=U.$ The last part of this theorem follows from Lemma 2.6 and the fact that $\rm X$ is also a $\rm \bigvee_w {\cal A}$-system. This completes the proof. \par\medskip
    
\noindent {\bf 4. An Ultralogic Unification}\medskip Assume for nonempty $\Sigma,$ that $\vert \Sigma \vert \leq \aleph_0.$ Let $\rm {\cal C}$ denote a set of (logic-system) corresponding finitary consequence operators each considered as defined on the language $\rm L.$ There exists a surjection $\rm f\colon \nat \to {\cal C}$ such that $\rm f(i)$ is one of the members of $\cal C$ and for each ${\rm C}\in {\cal C}$ there is some ${\rm j} \in \nat$ such that $\rm f(j) = C.$  For each $\rm i \in \nat$, let $\rm f(i) = C_i$ denote a consequence operator defined on L. As usual, for the next theorem, we use the boldface type convention (Herrmann, 1993, p. 21), and for the case $\cal C,$ $\underline {\cal C}$, will denote boldface type.  \par\medskip
{\bf Theorem 4.1.} {\it Let $\rm L$ and $\rm \{C_i \mid i \in \nat\}={\cal C}$ be defined as above. Suppose that every (non-empty) finite subset of $\rm {\cal C}$ is system consistent. \par
\indent \indent {\rm (i)} Then there exists a hyperfinite ultralogic $\cal U \in \Hyper {{\cal C}_{\bf f}}$ defined on the set of all internal subsets of $\Hyper {{\bf L}}$ such that ${\cal U} \not= \Hyper {{\bf U}},$ and an internal $W \subset \Hyper {{\bf L}}$ such that,  for each $\rm C_i \in {\cal C},$ 
$\Hyper {{\bf C}}_i(W) = {\cal U}(W)= W \not= \Hyper {{\bf L}},$ where ${\cal U}(W) \subset \Hyper {{\bf L}}.$\par\smallskip
\indent\indent {\rm (ii)}  For each internal $Y \subset \Hyper {{\bf L}},\ \bigcup \{\Hyper {{\bf C}}_i(Y) \mid i \in \nat\} \subset {\cal U}({Y})\subset \Hyper {{\bf L}}$. \par\smallskip
\indent\indent {\rm (iii)} If finite $\rm X \subset L,$ then $\bigcup\{
\Hyper{\bf C}_i({\bf X})\mid i \in \nat\} \subset {\cal U}({\bf X}),$ and if each member of $\cal C$ is a practical consequence operator, then $\bigcup\{
{\bf C}_i({\bf X})\mid i \in \nat\} \subset {\cal U}({\bf X}).$ \par\smallskip
\indent\indent {\rm (iv)} Let $\rm X\subset L$ and $\rm X \not= L$ be a {\rm C}-system for each $\rm C \in \cal C$. Then $\Hyper {{\bf X}}=
\Hyper{\bf C}_i(\Hyper {{\bf X}}) = {\cal U}(\Hyper {{\bf X}})\not= \Hyper {{\bf L}}$, for each $ i \in \nat.$ If $\rm X$ is finite ${\bf X} = \Hyper {{\bf C}_i}({\bf X})= {\cal U}({\bf X}),$ for each $i \in \nat.$ If for $\rm j \in \nat,$ $\rm C_j$ is a practical consequence operator, then  ${\bf X}= {\bf C}_j({{\bf X}}) = {\cal U}({\bf X})= {\cal U}({\bf X}).$}\par\smallskip
Proof. Let $\rm \langle {\cal C}_f, \lor_w, \land, I, U \rangle$ be the lattice of all finitary consequence operators defined on $\rm L$.  Consider this lattice, all of our intuitive consequence operators, our L  and all other defined objects as embedded into the Grundlagen Structure $\cal Y$ (Herrmann, 1993). Hence, they are embedded, in the usual manner, into the superstructure model, ${\cal M}= \langle {\cal N},\in,=\rangle,$ for all bounded formal expressions and this is further embedded into the superstructure $\cal Y$ that contains a nonstandard elementary extension $\Hyper {\cal M} =\langle \Hyper {{\cal N}}, \in,=\rangle$ of the embedded ${\cal M}$. Notice that from our identifications, any standard $\bf X \subset L$ has the property that $^{\sigma}{\bf X = X},$ and if $\bf X$ is finite, then $\Hyper {{\bf X}} = {\bf X}.$ Under our basic embedding, let $g\colon \nat \to \underline{\cal C}$ be a surjection in $\cal N$ that corresponds to $\rm f.$ Now consider the surjection $\hyper {g} \colon \hypernat \to \Hyper {\underline{\cal C}}.$ Let constant $a \in \nat$.  Under our special Grundlagen embedding procedures, $\Hyper {(g(a))} = \hyper {g(\hyper {a})} = \Hyper {g}(a)= \Hyper {{\bf C}_a}.$ Since $\hyper g$ is a surjection, an $a \in \nat$ corresponds to a member of ${}^\sigma \underline{\cal C}$ and conversely.  Thus, $\hyper {g}$ restricted to members of $^\sigma\nat = \nat$ yields the entire set ${}^\sigma \underline{\cal C}$.\par\smallskip 
Let nonempty $\rm K \subset \power {L}$ be the set of all $\rm X \not= L$ that if $\rm X \in K,$ then $\rm X$ is a C-system for each $\rm C \in {\cal C}.$ By Theorem 3.8, the definitions and the properties of the lattice structure on $\rm {\cal C}_f,$ for clarity, the unsimplified and redundantly expressed bounded sentences\par 
$$\forall x((x \not= \emptyset) \land(x \in {\cal F}(\nat)) \to \exists y \exists w_1((y \in {\cal C}_{\bf f}) \land ( y \not= {\bf U}) \land (w_1 \in\power {\bf L}) \land $$ $$
(\forall z_1\forall v_1 \forall v_2 ((v_1 \in x) \land (v_2 \in {\bf K})\land (v_1 \in \nat)\land (z_1 \in x)\land(z_1 \in \nat) \to (g(z_1)(v_2) = $$ $$y(v_2)=v_2 \subset {\bf L})\land(y(v_2) \not= {\bf L})))\land 
(\forall v( (v \in x)\land (v \in \nat) \to $$ $$g(v)({w_1}) = y({w_1})= w_1 \not= {\bf L}))\land  \forall z((z \in x) \land (z \in \nat) \to  ((g(z) \leq y) \land $$ $$\forall w((w \in {\bf {\cal C}_f}) \land \forall z_1((z_1 \in x)\land (z_1 \in \nat) \land (g(z_1)\leq w))\to (y \leq w)))))),\eqno (4.1.1)$$ 
$$\forall x\forall y((x \in {\bf {\cal C}_f}) \land (y \in {\bf {\cal C}_f})\to ( (y \leq x) \iff $$ $$\forall w(( w \in \power {{\bf L}}) \to (y(w) \subset x(w))))), \eqno (4.1.2)$$ 
hold in $\cal M$. Hence, they hold under *-transfer in $\Hyper {\cal M}$ for objects in $\Hyper {\cal N}.$ [Note: It is usually assumed that formal statements such as 4.1.1 and 4.1.2 can be made within a formal first-order language rather than expressing them explicitly.] \par\smallskip  
The set $\Hyper{{\bf {\cal C}_f}}$ is a collection of hyperfinite consequence operators defined on the internal subsets of $\Hyper {{\bf L}}.$ Let infinite $\lambda \in \hypernat - \nat.$ Then $\Hyper g[[0,\lambda]]$ is a hyperfinite subset of $\Hyper {{\underline{\cal C}}} \subset \Hyper {{\bf {\cal C}_f}}.$ Hence, from *-transformed sentences (4.1.1) and (4.1.2), there exists some hyperfinite ${\cal U} \in \Hyper {{\bf {\cal C}_f}}$ defined on the set of all internal $Y\subset \Hyper {{\bf L}}$ with the properties that $\Hyper g(i)(Y) \subset {\cal U}(Y),$ for each $i \in [0, \lambda],$ and, in particular for $i \in \nat.$ Hence, $\bigcup \{\Hyper {{\bf C}}_i(Y) \mid i \in \nat\} \subset {\cal U}({Y}).$ 
 Further, there exists an internal $W \subset \Hyper {{\bf L}}$ such that, for each $i \in [0, \lambda],$ $\hyper g(i)(W) = {\cal U}({W})= W \not= \Hyper {{\bf L}}.$ In particular, $\hyper g(i)(W) = \Hyper {{\bf C}_i}(W) = {\cal U}(W),$ for each $i \in \nat.$ Since $ {\cal U}({W}) \not= \Hyper {{\bf L}},$ then ${\cal U} \not= \Hyper {{\bf U}}.$  Let finite $\rm X \subset L$. Then  due to our embedding procedures, $\Hyper {({\bf C}_i}({\bf X}))= \Hyper {{\bf C}_i}(\Hyper {{\bf X}}) = \Hyper {{\bf C}_i}({{\bf X}})\subset {\cal U}({\bf X})\not= \Hyper {{\bf L}}.$  Hence, $\bigcup\{\Hyper {{\bf C}}_i({\bf X})\mid i \in \nat\} \subset {\cal U}({\bf X}).$  [Note: in proofs such as this and to avoided confusion, I often, at first, use the notation $\Hyper {({\bf C}_i}({\bf X}))$ to indicate the value (or name) of the result of applying $\hyper {}$ to an object in ${\cal N}$ which is a set such as ${\bf C}_i({\bf X})$ that contains additional operator notation. From a technical viewpoint, $\Hyper {({\bf C}_i}({\bf X}))= \Hyper \{{\bf C}_i({\bf X})\} =\{\Hyper {{\bf C}_i}({\bf X})\}$ and  $\Hyper {{\bf C}_i}({\bf X})$ is the ``name'' for the set under the mapping *. But using this procedure, there is confusion as to whether $\Hyper {({\bf C}_i}({\bf X}))$ denotes the entire set or denotes the operator $\Hyper {{\bf C}_i}$ applied to ${\bf X}.$ In these proofs, $\Hyper {{\bf C}_i}$ always denotes the operator $\Hyper {{\bf C}_i}$ applied to internal subsets of $\Hyper {{\bf L}}.$] If $\rm C_i$ is a practical consequence operator, then $\rm C_i(X)$ is a finite set. Hence $\Hyper {{\bf C}_i}({\bf X}) = {\bf C}_i(\bf X).$ (iv) follows by *-transfer and this completes the proof.\par\smallskip
In Herrmann (2001a, b), the set $\rm \{\rm S^V_{N_j}\mid j \in \nat\}$ is the refined set of all relativized axiomless science-community consequence operators defined on a language $\Lambda$ and they are used to unify, in a restricted manner, physical theory behavior. Moreover, for sequentially presented $ \{\rm S^V_{N_j}\mid j \in \nat\},$ $\rm 1 \leq \vert \{\rm S^V_{N_j}\mid j \in \nat\}\vert \leq \aleph_0.$\par\medskip
{\bf Corollary 4.1.1.} \par\smallskip {\it 
\indent\indent {\rm (i)} There exists a hyperfinite ultralogic ${\cal U} \in 
 \Hyper {{\cal C}_{\bf f}}$
 that is defined on all internal $Y \subset \hyper {{\bf \Lambda}}$ such that ${\cal U} \not= \Hyper {{\bf U}},$ and, for each $i \in \nat,\ 
\Hyper {{\bf S^V_{\bf N}}_i}(\emptyset) = {\cal U}(\emptyset)= \emptyset.$\par  \indent\indent {\rm (ii)} For each  internal $Y \subset \hyper {{\bf \Lambda}},  \bigcup\{\Hyper {{\bf S^V_{\bf N}}_i}({ Y})\mid i \in \nat\} \subset {\cal U}({ Y})\subset \hyper {{\bf \Lambda}}.$\par \indent\indent {\rm (iii)} If finite $\rm X \subset \Lambda,$ then $\bigcup\{
\Hyper {{\bf S^V_{\bf N}}_i}({\bf X})\mid i \in \nat\} \subset {\cal U}({\bf X}),$ and if each member of $\rm \{ S^V_{N_i}\mid i \in \nat\}$ is a practical consequence operator, then $ \bigcup\{
{\bf S^V_{\bf N}}_i({\bf X})\mid i \in \nat\} \subset {\cal U}({\bf X})$.\par \indent\indent {\rm (iv)} Let $\rm X \subset \Lambda,\ X \not= \Lambda$ be a {\rm C}-system for each $\rm C \in \{{\rm S^V_{N}}_i({X})\mid i \in \nat\}.$ Then $\Hyper {{\bf X}} = \Hyper {{\bf S^V_{\bf N}}_i}(\Hyper {{\bf X}})= {\cal U}(\Hyper {{\bf X}})\not= \Hyper {{\bf \Lambda}}$, for such each $i \in \nat.$ If $\rm X$ is finite, then  $\Hyper {{\bf S^V_{\bf N}}_i}({{\bf X}}) = {\bf X}= {\cal U}({\bf X}),$ for each $i \in \nat.$ If for $\rm j\in \nat$, $\rm S^V_{N_j}$ is a practical consequence operator, then  ${{\bf S^V_{\bf N}}_j}({{\bf X}}) = {\bf X}= {\cal U}({\bf X}).$} \par\medskip
\noindent {\bf 5. Further Applications}\medskip

If the $\cal C$ in the hypotheses of Theorem 4.1 is restricted to a set of practical consequence operators, each defined on a finite language $\rm L^f$, then it follows that the hyperfinite $\cal U$ corresponds to a hyperfinite logic-system ${\cal S}$ that is *-effectively *-generated. If the effective notion is not required, then, in general, the ultralogic  $\cal U$ corresponds to a *-logic-system. 
Although Theorem 4.1 and Corollary 4.1.1 are mainly concerned with the original set of consequence operators $\cal C$ and $\rm \{S^V_{N_i}({ X})\mid i \in \nat\}$, when $\vert \rm \{S^V_{N_i}({ X})\mid i \in \nat\}\vert = \aleph_0,$ it is also significant for applications that $\cal U$ unifies each ``ultranatural relativized theory'' $\hyper g(j),\ j \in [0,\lambda] - \nat.$ It follows that for each $j \in [0,\lambda] -\nat,\ \hyper g(j)(\emptyset) = {\cal U}(\emptyset) = \emptyset$ and for internal $Y \subset \Hyper {\bf \Lambda},\ \hyper g(j)(Y) \subset {\cal U}(W).$ This also applies to the unrelativized case with the  modifications that $\emptyset$ is replaced with $W$ and each relativized consequence operator $\rm S^V_{N_i}$ is replaced with the physical theory consequence operator $\rm S_{N_i}$. Also note that $\rm S_{N_i}$ and $\rm S^V_{N_i}$ are usually considered practical consequence operators.\par\smallskip
Depending upon the set $\cal C$ of consequence operators employed, there are usually many $\rm X \subset L, \ X \not= L$ such that $\rm X$ is a C-system for each $\rm C \in \cal C.$ For example, we assumed in Herrmann (2001a, b) that there are two 1-ary relations for the science-community logic-systems. One of these contains the logical axioms and the other contains a set of physical axioms; a set of natural laws. Let $\rm \{S'_{N_i}\mid i \in \nat\}$ be the set of science-community corresponding consequence operators relativized so as to remove the set of logical theorems. Each member of a properly stated set of natural laws, $\rm N_j,$ used to generate the consequence operators $\rm \{S'_{N_i}\mid i \in \nat\}$ should be a C-system for each 
member of $\rm \{S'_{N_i}\mid i \in \nat\}$. As mentioned, the physical theories being considered here are not theories that produce new ``natural laws.'' The argument that the Einstein-Hilbert equations characterize gravitation fields, in general, leads to the acceptance by many science-communities
of these equations as a ``natural law'' that is then applied to actual physical objects. Newton's Second Law of motion is a statement about the notion of inertia within our universe. It can now be derived from basic laboratory observation and has been shown to hold for other physical models distinct from its standard usage (Herrmann, 1998). The logic-systems that generate the members of $\rm \{S'_{N_i}\mid i \in \nat\}$ have as a 1-ary relation a set of natural laws. Then one takes a set of specific physical hypotheses $\rm X$ that describes the behavior of a natural-system and applies the logic-system to $\rm X.$ This gives a statement as to how these natural laws affect, if at all, the behavior being described by $\rm X$. It is this approach that implies that each properly described $\rm N_j \not= L$ is a C-system for each 
$\rm C \in  \{S_{N_i}\mid i \in \nat\}$. Applying Theorem 4.1 to $\rm {\cal C} =\{S'_{N_i}\mid i \in \nat\},$ where $\rm \bigcup \{{N}_i\mid i \in \nat\}\subset K,$ leads to a result exactly like Corollary 4.1.1 where results (i), (iii) and (iv), applied to members of $\rm \{N_i\mid i \in \nat\},$ are particularly significant. This result is stated as the {\it Abstract} for this paper.\par\smallskip
At any moment in human history, one can assume, due to the parameters present, that there is, at the least, a denumerable set of science-community logic-systems or that there exist only a finite collection of practical logic-systems defined on finite $\rm L^f.$ The corresponding set $\rm {\cal C}^f = \{C^f_i\mid i = 1,\ldots, n \}\subset {\cal C}^f_f$ of practical consequence operators would tend to vary in cardinality at different moments in human history. For the corresponding finite set of practical consequence operators, by Theorem 2.10, there is a standard (least upper bound) practical consequence operator $\rm {\cal U}$, and hence ``the best'' practical logic-system, that unifies such a finite set. The following result is the interpretation of Theorem 4.1 for such a finite set of practical consequence operators.  \par\medskip 
{\bf Theorem 5.1.} {\it Let $\rm L^f$ and $\rm {\cal C}^f$ be defined as above. Suppose that $\rm {\cal C}^f$ is system consistent. \par
\indent \indent {\rm (i)} Then there exists a practical consequence operator $\rm {\cal U}_1 \in {\cal C}_f^f$ defined on the set of all subsets of $\rm L^f$ such that $\rm {\cal U}_1 \not=  U,$ and a $\rm W \subset L$ such that,  for each $\rm C_i^f \in {\cal C}^f,$ 
$\rm C_i^f(W) = {\cal U}_1(W)= W \not= L^f,$ where $\rm {\cal U}_1(W) \subset  L^f.$\par\smallskip
\indent\indent {\rm (ii)}  For each $\rm X \subset L^f,\ \bigcup \{C_i^f(X) \mid i=1,\cdots,n\} \subset {\cal U}_1(X)\subset L^f$ and ${\cal U}_1$ is the least upper bound in  $\rm \langle {\cal C}^f_f, \lor_w, \land, I, U \rangle$ for $\rm {\cal C}^f.$  \par\smallskip
\indent\indent {\rm (iii)} Let $\rm X\subset L^f$ and $\rm X \not= L^f$ be a ${\rm C_i^f}$-system for each $\rm C_i^f \in {\cal C}^f$. Then $\rm X=
C_i^f(X) = {\cal U}_1(X)\not= L^f$, $ \rm i=1,\cdots,n.$} \par\medskip

 Letting finite $\rm {\cal C}^f$ contain practical consequence operators either of the type $\rm S_{N_i},$  $\rm S^V_{N_i}$ or $\rm S'_{N_i},$ exclusively, then $\rm {\cal U}_1$ would have the appropriate additional properties and would generate a practical logic-system. Corollary 2.10.1 and Theorem 3.8 yield a more general unification $\rm \bigvee_w {\cal A},\ {\cal A} \subset {\cal C}_f,$ as represented by a least upper bound in $\rm \langle {\cal C}_f,\lor_w, I, U \rangle,$ with the same properties as stated in Theorem 5.1. Thus depending upon how physical theories are presented and assuming system consistency, there are nontrivial standard unifications for such physical theories. Assuming that $\vert {\cal A} \vert = \aleph_0,$ and that $\cal A$ is system consistent, then the Corollary 2.10.1 unification $\rm \bigvee_w {\cal A}$ corresponds to a (nontrivial) nonstandard ultralogic unification $\Hyper {\bigvee_{\bf w} {\cal A}}$ with all of the same stated properties as those of the $\cal U.$ However, $\Hyper {\bigvee_{\bf w} {\cal A}}$ and $\cal U$ have one significant difference. The ultralogic $\cal U$ is, with respect to internal subsets of $\Hyper {\bf L},$ a ``least upper bound''  of a hyperfinite collection, where as $\Hyper {\bigvee_{\bf w} {\cal A}}$ need not have this additional hyperfinite property. Further, system consistency is used only so that one statement in Theorem 4.1, Corollary 4.2, Theorem 5.1 and this paragraph will hold. This one fact is that each of the standard unifications of a collection ${\cal A}\subset {\cal C}_f$ is not the same as the upper unit if and only if the ${\cal A}$ is system consistent. Further, if an $\rm X\subset L^f$ [resp. $\rm X \subset L$] is ${\cal U}_1$-consistent [resp $\rm \bigvee_w {\cal A}$-consistent], then $\rm X$ is C-consistent for each $\rm C \in {\cal C}^f$ [resp. $\rm C \in {\cal A}$].\par\smallskip 
For General Intelligent Design Theory, the ultralogic ${\cal U},\ {\cal U}_1$ or $\Hyper {\bigvee_{\bf w} {\cal A}}$ can replace the $\rm {\cal S}_W$ and $\Pi$ discussed in Herrmann (2001a, b) and can, obviously, be interpreted as an intelligence that designs and controls the combined behavior exhibited by members of $\rm {\cal C} = \{S'_{N_i}\mid i \in \nat\},$  as they are simultaneously applied to a natural-system. For denumerable $\cal C,$ it is also significant that the ultralogics $\{ \Hyper {g(i)} \mid i \in [0,\lambda] - \nat \}$ [resp. $D \in \Hyper {{\cal C}} - {\cal C}$] represent ultranatural theories relative to ultranatural laws contained in $\Hyper {\{}{\bf N}_i\mid i \in \nat\}$ and these ultralogics are unified by $\cal U$ [resp. $\Hyper {\bigvee_{\bf w} {\cal A}}$]. These ultranatural laws would lead to various ultranatural events as being required constituents associated with event sequences. Further, from the definition of refined theories as discussed in Herrmann (2001a, b), $\Hyper {\{}{\bf N}_i\mid i \in \nat\}$ contains descriptions for the ultrasubparticles. This is a second prediction that these various ``ultra'' objects can be assumed to rationally exist, where they were first discovered by other methods that appear in Herrmann (1993). \par\medskip  
{\bf Remark 5.2.} I mentioned that the nonstandard results obtained in section 4.1 are established by means of the most trivial methods used in Robinson-styled nonstandard analysis.\par \medskip

\centerline{\bf References}\par\medskip
\id{D}ziobiak, W. (1981), ``The lattice of strengthenings of a strongly finite consequence operation,'' {\it Studia Logica} 40(2):177-193.\smallskip
\id{H}errmann, R. A. (2001a). ``Hyperfinite and Standard Unifications for Physical Theories,'' {\it Intern. J. of Math. Math. Sci.}, 28(2):93-102.\smallskip\smallskip
\id{H}errmann, R. A. (2001b). ``Standard and Hyperfinite Unifications for Physical Theories,''
http://www.arXiv.org/abs/physics/0105012\smallskip\smallskip
\id{H}errmann, Robert A. (2001c), ``Ultralogics and probability models,'' {\it Intern. J. of Math. Math. Sci.}, 27(5):321-325.\smallskip
\id{H}errmann, R. A. (2001d), ``Probability Models and Ultralogics,''\hfil\break
http://www.arXiv.org/abs/quant-ph/0112037\smallskip
\id{H}errmann, Robert A. (1998), ``Newton's second law of motion holds in normed linear spaces,'' {\it Far East J. of Appl. Math.} 2(3):183-190. \smallskip

\id{H}errmann, Robert A. (1993),  {\it The Theory of Ultralogics.} \hfil\break http://www.arXiv.org/abs/math.GM/9903081 and/9903082\smallskip

\id{H}errmann, Robert A. (1987), ``Nonstandard consequence operators''. {\it Kobe Journal of  Mathematics} 4:1-14. http://www.arXiv.org/abs/math.LO/9911204\smallskip
\id{\L}o\'s, J. and R. Suszko, (1958), ``Remarks on sentential logics,''
{\it Indagationes Mathematicae} 20:177-183.\smallskip

\id{T}arski,  Alfred. (1956), {\it Logic, Semantics, Metamathematics; papers from 1923 - 1938},  Translated by J. H. Woodger$.$ Oxford: Clarendon Press.
\smallskip
\id{W}\'ojcicki, R. (1973), ``On matrix representations of consequence operators on \L ukasiewicz's Sentential Calculi,'' {\it Zeitschi. f. math. Logik und Grundlagen d. Math.,} 19:239-247.
\smallskip
\id{W}\'ojcicki, R. (1970), ``Some remarks on the consequence operation in sentential logics,'' {\it Fundamenta Mathematicae}, 68:269-279.\smallskip

\end